# Black hole evaporation without an event horizon

James M. Bardeen
Physics Department, University of Washington
Seattle, Washington 98195-1560
email: bardeen@uw.edu

**Abstract**. A reformulation of the calculation of the semi-classical energy-momentum tensor on a Schwarzschild background, the Bousso covariant entropy bound, and the ER=EPR conjecture of Maldacena and Susskind taken together suggest a scenario for the evaporation of a large spherically symmetric black hole formed by gravitational collapse in which 1) the classical $r = 0$ singularity is replaced by an initially small non-singular core inside an inner apparent horizon, 2) the radius of the core grows with time due to the increasing entanglement between Hawking radiation quanta outside the black hole and Hawking partner quanta in the core contributing to the quantum back-reaction, and 3) by the Page time the trapped surfaces disappear, and all quantum information stored in the interior is free to escape. The scenario preserves unitarity without any need for a "firewall" in the vicinity of the outer apparent horizon. Qbits in the Hawking radiation are never mutually entangled, and their number never exceeds the Bekenstein-Hawking entropy of the black hole. The quantum back-reaction, while it must be very large in the deep interior of the black hole, can be described semi-classically in the vicinity of the outer apparent horizon up until close to the Page time. An explicit toy model for the metric in the interior of the black hole in this scenario, and how its associated energy-momentum tensor can be continued to the exterior in a semi-classical approximation, is discussed.

## I. INTRODUCTION

The original derivation of Hawking radiation from black holes[1] and the prediction of its essentially thermal character at the Hawking temperature

$$T_\text{H} = \frac{\kappa}{2\pi} \left(\hbar = G = c = 1\right) = \frac{\kappa m_\text{p}^2}{2\pi} \left(G = c = 1, \hbar = m_\text{p}^2\right), \quad (1.1)$$

in which the surface gravity of the horizon $\kappa = 1/(4M)$ for a Schwarzschild black hole of mass $M$, was based on semi-classical effective field theory. While the Planck mass $m_\text{p}$ is often set equal to one, I will not do so, in order to emphasize the smallness of quantum corrections for a large astrophysical black hole, for which $\left(m_\text{p}/M\right)^2 < 10^{-76}$. The quantum fields are considered test fields in the curved spacetime of a classical background geometry, a solution of the classical Einstein equations. Calculations of the energy flux in outgoing radiation[2] for various types of fields on a Schwarzschild background, taking into account partial transmission of field modes through a potential barrier centered around $r = 3M$, were followed by calculations of the renormalized expectation value of the complete quantum energy-momentum tensor outside the classical event horizon, $r > 2M$, using a framework established by Christensen and Fulling[3]. The usual assumptions are that the energy-momentum tensor is time-independent, satisfies local energy-momentum conservation, and has a trace determined solely by the conformal anomaly, not only for the fields that are classically conformally invariant, such as a massless scalar field and the electromagnetic field, but also for the gravitational field.

The calculations require specifying a "vacuum" state for each field. Results were first obtained for the Hartle-Hawking (HH) state[4], which assumes thermal equilibrium (and therefore no net energy flux), with the outgoing Hawking radiation balanced by incoming radiation from an external heat bath at the Hawking temperature. However, "real" black holes are formed by gravitational collapse, and the natural vacuum state is the absence of particles at past null infinity in the asymptotically Minkowski spacetime. Unruh[5] noted that this vacuum could be realized in the $r > 2M$ region of the maximally extended Schwarzschild spacetime by demanding the absence of incoming radiation at both past null infinity and the past horizon, plus regularity of the energy-momentum tensor on the future event horizon (the only one that actually exists in a collapse scenario) in the frame of a freely falling observer. In this paper the primary focus is on the Unruh vacuum.

Precise calculations are somewhat easier for the HH vacuum, with complete numerical results for a massless scalar field on a Schwarzschild background first obtained by Howard and Candelas[6], improving on approximate analytic estimates of Page[7]. For the Unruh vacuum, numerical results were obtained by Elster[8] for a massless scalar field and by Jensen, et al[9] for the electromagnetic field. In Part II of the paper I will discuss the physical interpretation of the Unruh vacuum results in some detail, and the Hartle-Hawking vacuum results more briefly. To facilitate this I first decompose the energy-momentum tensor into a "conformal anomaly" part and a traceless "regular" part, each separately conserved. The regular part is further split into physically motivated interacting pieces. This decomposition is quite different from the traditional one of Christensen and Fulling. Simple analytic expressions that approximate the known numerical results are formulated. For the scalar field case, they are similar to and numerically equivalent to those of Visser[10]. In my decompositions, the anomaly part strongly dominates the regular part in the vicinity of the horizon for spin 1 fields, and even more so for spin 2 fields.

My formulation clearly supports the Unruh[11] physical interpretation of how the outgoing Hawking radiation is generated. Splitting the net energy flux into an outgoing part and an ingoing part shows that the outgoing part originates somewhat outside the horizon, in the general vicinity of the potential barrier in the mode equations. The semi-classical results are *not* consistent with modeling the Hawking radiation as arising from pair creation within a Planck length of the horizon, as in Parikh and Wilczek[12]. If the Hawking radiation came from that close to the horizon a freely falling observer crossing the horizon would see an enormous spike in the energy density there. The Unruh picture is that the Hawking radiation is generated non-locally by the distortion of the vacuum modes of the quantum fields as they propagate in the geometry of the black hole spacetime. The Hawking radiation can reasonably be given a particle interpretation only well outside the black hole. The Hawking quanta must be entangled with quantum degrees of freedom inside the black hole, which can be described as Hawking "partner" quanta well inside the black hole horizon. The quantum information in the Hawking partners accumulates steadily inside the black hole as the black hole evaporates, unless some mechanism exists to transfer this information out across the horizon in the later Hawking radiation. No such mechanism is apparent semi-classically, since local quantum field theories can only propagate information in or on the future light cone and there is no indication of concentration of radiation near the horizon that could carry significant quantum information.

In Section III, I examine the first-order back-reaction on the metric, using the expectation value of the quantum energy-momentum tensor as a source in the spherically symmetric Einstein equations. The calculations are done in advanced Eddington-Finkelstein coordinates, in which the time coordinate $v$ is constant on ingoing radial null geodesics. The area of the 2-spheres defines the circumferential radius $r$, from which a mass function $m(v,r)$ is defined by

$$\nabla^\mu r \nabla_\mu r \equiv 1 - 2m/r. \tag{1.2}$$

An apparent horizon is located wherever $r = 2m$, at which the gradient of $r$ is null, and (for a *black* hole) is on the boundary of a region of trapped surfaces, where the gradient of $r$ is timelike and $r$ decreases toward the future. The mass function can be interpreted as the total energy inside the 2-sphere labeled by $r$ and in an asymptotically flat space time is the ADM mass at spatial infinity and the Bondi mass $M$ at future null infinity. What one finds from the usual form of the semi-classical energy-momentum tensor (see Bardeen[13]) is that the geometry retains to a good approximation the Schwarzschild form, with

$$(\partial m/\partial v)_r = 4\pi r^2 \langle T_v^r \rangle = -L_\mathrm{H} \tag{1.3}$$

at all $r$, where $L_H$ is the Hawking luminosity. The original calculations of Page[2], recently reviewed in Ref. [14], determined the spin-dependent numerical coefficients $k_s$ defined by

$$L_H = \frac{4\pi}{245760\pi^2} \frac{m_\mathrm{P}^2}{M^2} \sum_s k_s = 4\pi M^2 \sigma T_\mathrm{H}^4 \sum_s k_s, \tag{1.4}$$

where $\sigma = \pi^2/(60 m_\mathrm{P}^6)$ is the Stefan-Boltzmann constant. For a solar mass or larger black hole only spin 1 photons ($k_1 = 6.49$) and spin 2 gravitons ($k_2 = 0.742$) are expected to contribute, while a hypothetical massless scalar would have $k_0 = 14.26$.

Naively, the standard semi-classical analysis should remain valid and the evaporation should continue until the black hole mass becomes roughly the order of the Planck mass. This is based on a key assumption, that the trace of the energy-momentum is entirely due to the conformal anomaly arising from the renormalization of massless fields. However, classical general relativity is not conformally invariant, and there could be an additional non-anomalous quantum gravity contribution to the effective $T_\alpha^\alpha$. It is this possibility that is the basis of the model of enhanced quantum back-reaction developed later in the paper.

The standard semi-classical scenario is very difficult to reconcile with the unitarity expected of quantum field theories. The number of quantum degrees of freedom associated with a black hole of mass $M$ is thought to have an upper limit equal to the Bekenstein-Hawking entropy

$$S_{\text{BH}} = 4\pi \left(\frac{M}{m_{\text{p}}}\right)^2, \tag{1.5}$$

which is the coarse-grained thermodynamic entropy of the black hole. As discussed in detail in a recent review of Page (2013), the quantum information inside the black hole saturates the Bekenstein-Hawking bound at the Page time, defined as the time when the von Neumann entropy of the emitted Hawking radiation equals $S_{\text{BH}}$, i.e., when $S_{\text{BH}}$ has decreased to roughly one-half of its initial value. Unitary evolution from an initial pure state with zero von Neumann entropy requires that somehow the interior quantum information *must* be able to escape from the black hole if the evaporation continues past the Page time, so that the von Neumann entropy of the Hawking radiation can be reduced to zero by the time the black hole evaporates completely or almost zero if there is a Planck scale remnant. The AdS-CFT correspondence[15] has usually been interpreted as implying that a unitary S-matrix describes the complete process of formation and evaporation of the black hole. Whether the spacetime is asymptotically AdS or flat should be immaterial if the recurrence time of the AdS spacetime is longer than the evaporation time of the black hole. The problem is that entanglement of late Hawking quanta, emitted after the Page time, with early Hawking quanta is necessary for unitarity, while entanglement of the late Hawking quanta with their partners inside the black hole is necessary if the quantum fields are to be in a near vacuum state in the vicinity of the horizon as measured by a freely-falling observer and as required from the local flatness expected from general relativity. But quantum mechanics requires monogamy of entanglement. There is also the issue of how quantum information carried into the black hole in the initial collapse and by the early Hawking partners can be transferred outside the black hole without violating the locality expected of semi-classical field theories.

The idea of complementarity was introduced by Susskind, et al[16] as a way to deal with this conflict. They argue that an external observer ("Bob") and a freely falling observer ("Alice") can have very different, but complementary, descriptions of the physics of black hole evaporation. According to Bob the black hole is a quantum system in a thermal state with $S_{\text{BH}}$ degrees of freedom, at least after a fairly short "scrambling time" of order $M \log(S_{\text{BH}})$. It is assumed that, from Bob's point of view, quantum information does not disappear into the black hole, but is preserved in a "stretched horizon" of Planck scale thickness just outside the event horizon. The quantum information comes out in the Hawking radiation more or less according to the Page scenario. On the other hand, Alice sees approximate vacuum crossing the horizon and can measure the quantum information in the Hawking partners inside the black hole that have not been absorbed by the central singularity She cannot detect cloning of quantum information into the stretched horizon, another potential violation of quantum mechanics, or the entanglement of late Hawking quanta with the early Hawking quanta.

The complementarity scenario does not make much sense in the context of semi-classical theory, in which the quantum fields must propagate causally on what should be, to an extraordinarily good approximation in the vicinity of the horizon for a large black hole, a classical GR background. The location of an actual event horizon is not determined by local physics; it depends on the entire future history of the interaction of the black hole with its surroundings, since it is defined as the boundary of the past of future null infinity. Even an apparent horizon requires knowledge of the geometry over an entire sphere. It is hard to see how local Planck scale physics would distinguish a black hole apparent horizon from the Rindler horizon associated with uniformly accelerating observers in Minkowski spacetime. Rindler observers see a heat bath at a temperature related to their acceleration, but no energy flux corresponding to Hawking radiation. Rindler observers cannot retrieve quantum information that disappears behind the Rindler horizon. Furthermore, the apparent horizon typically starts as a spacelike hypersurface as the black hole forms, and is temporarily spacelike if a massive shell collapses onto the black hole at some later time, so not all of the quantum information in the collapsing star and the Hawking partners can reliably propagate causally along the apparent horizon until it can be released in the Hawking radiation at some much later time. The "scrambling time", either classically or semi-classically, is the time for a disturbance in the "stretched horizon" to dissipate by propagating *away* from the horizon, returning the quantum fields near the horizon to the original Unruh vacuum state.

The recent controversy was set off by the paper of Almheiri, et al[17] (AMPS), though similar ideas had been floated by others previously. AMPS argue that at least after the Page time, when unitarity requires entanglement of late Hawking quanta with early Hawking quanta, the quantum fields cannot be close to a vacuum state as seen by a freely falling observer crossing the horizon. Instead, they must be in a highly excited state, containing super-Planckian energies, a "firewall" that would vaporize the observer. A flood of responses ensued, many questioning the need for firewalls, but still invoking major departures from general relativity and/or conventional quantum field theory. Later papers by members of the AMPS collaboration responded to critics[18] and offered additional arguments for why firewalls seem to be necessary, perhaps even before the Page time[19]. A problem with the firewall argument is that while some quantum states may have firewalls, it is hard to see how firewalls could possibly arise from what is the vacuum state prior to the formation of the black hole. This point has also been made by Page[20] and Freivogel[21].

The firewall argument depends critically on the assumption that Hawking radiation continues past the Page time, with a final state entirely or almost entirely consisting of Hawking quanta. What I present is a quite different scenario, in which quantum back-reaction grows and destroys all trapped surfaces at or before the Page time. This enhanced quantum back-reaction is definitely not consistent with the standard semi-classical calculations of the effective energy-momentum tensor. However, there are reasons to suspect that the standard calculations may be missing an important contribution from quantum fluctuations of the gravitational field, a contribution driven by the increasing amount of entanglement between the Hawking quanta outside the black hole and their partners inside the black hole. If the ER=EPR hypothesis of Maldacena and Susskind[22] is correct, and each Hawking quantum is connected by a microscopic Einstein-Rosen bridge to its partner inside the black hole, this could have a substantial effect on the macroscopic black hole geometry by the Page time. My scenario postulates a non-singular inner core to the black hole, inside an inner apparent horizon, containing the quantum information from the initial collapsing star and the accumulated Hawking partners. Applying the Bousso covariant entropy bound[23] at or just outside the inner apparent horizon implies a steady increase in the radius of the inner apparent horizon as the von Neumann entropy in the core increases, associated with a growing non-anomalous contribution to the trace of the energy-momentum tensor. By the Page time the core overtakes the outer apparent horizon, eliminating all trapped surfaces. The number of Einstein-Rosen bridges stretching between the core and the exterior of the black hole is always roughly the area of the *inner* apparent horizon in Planck units. The black hole has no event horizon, as argued on general grounds by Hawking[24], and all trapped quantum information is eventually able to escape.

Some groundwork for this scenario is laid in Section III, where I consider the back-reaction due to the semi-classical conformal anomaly, which can with some confidence be extrapolated to deep inside the black hole. The result is that the quantum back-reaction on the geometry becomes large at a radius large compared to the Planck radius (though the spacetime curvature is reaching the Planck scale). A much more speculative extrapolation of the corrected metric to still smaller radii suggests the possibility of a non-singular core to the black hole, with an inner apparent horizon radius the order of the Planck length.

An explicit model for the black hole interior, not completely realistic, but analytically simple, and having the right qualitative properties for the proposed scenario, is discussed in Section IV. An ansatz for the metric inside the black hole, an explicit function of Schwarzschild radius $r$ and advanced time $v$, is adopted and the corresponding energy-momentum tensor is calculated from the Einstein equations. The time dependence is implemented through a black hole mass function $M(v)$, decreasing in accord with the Page luminosity formula, and a function $a(v)$, to a good approximation the radius of the inner event horizon, which increases in accord with the Bousso entropy bound. I show how, as long as the metric is close to Schwarzschild in the vicinity of the outer apparent horizon, an exterior energy-momentum tensor of a form similar to that of Section II can be matched smoothly to the interior. The quantum back-reaction in the model is much larger than in standard semi-classical theory, and the curvature in the core is well below the Planck scale and decreases with time. No attempt is made to follow the disappearance of apparent horizons in any detail.

Finally, Section V summarizes the case for this scenario and contains some discussion of its broader implications. I argue that, while a large black hole radiates like a blackbody, it should not be thought of as a quantum system in a thermal state. It is the the *inner* apparent horizon boundary of the black hole *core* whose area is related to the von Neumann entropy of a quantum system. The Bekenstein-Hawking entropy of the black hole associated with the area of the outer apparent horizon is only an upper limit to the von Neumann entropy. If my scenario is correct, a black hole has a finite lifetime close to the Page time. A potential observational test of these ideas would be to observe the "explosions" of relatively small primordial black holes, since I would expect a considerably softer spectrum of particles and radiation than that predicted by the conventional scenario[25] of evaporation down to Planck scale.

## II. THE SEMI-CLASSICAL ENERGY-MOMENTUM TENSOR OUTSIDE THE SCHWARZSCHILD HORIZON

In the standard semi-classical approximation the effective energy-momentum tensor of quantum fields is first-order in $\hbar$, calculated on a fixed classical background geometry, taken here to be the spherically symmetric Schwarzschild geometry. At $r > 2M$ it is convenient to work with physical components of the tensor as projected onto the orthonormal frames of static observers, observers whose world lines remain at constant Schwarzschild radius r, with $4\pi r^2$ the area of a two-sphere, and the time coordinate $t$ is the usual Schwarzschild time coordinate labeling hypersurfaces orthogonal to these static world lines.

Due to the spherical symmetry there are only four independent components of the energy-momentum tensor, an energy density $E = -T_t^t$, an energy flux/momentum density $F = -T_t^r/(1-2M/r)$, a radial stress $P_r = T_r^r$, and a transverse stress $P_t = T_\theta^\theta = T_\varphi^\varphi$. Outside the horizon any disturbances not protected by global conservation laws dissipate by some combination of radiation out to future null infinity or inward across the horizon to the black hole interior. After some transient behavior associated with black hole formation, on a time scale $\sim M$ (the scrambling time), the expectation value of the energy-momentum tensor of a quantum field, it is assumed, should become stationary to first-order in $\hbar$. With time derivatives set to zero, the local energy-momentum conservation equations become

$$r^2 T_{t;\beta}^\beta = \partial_r \left[ r^2 (1-2M/r) F \right] = 0 \tag{2.1}$$

and

$$rT_{r;\beta}^\beta = (E + P_r)\frac{M/r}{1-2M/r} + \frac{1}{r}\partial_r(r^2 P_r) - 2P_t = 0. \tag{2.2}$$

Since the Hawking luminosity $L_H = \lim_{r\to\infty}(4\pi r^2 F)$, we see from Eqs. (1.4) and (2.1) that the contribution from spin $s$ to the net energy flux at finite $r$ is

$$F_s = \frac{k_s}{245760}\frac{m_P^2}{M^2}\frac{1}{r^2}\frac{1}{1-2M/r} = k_s \sigma T_H^4 \frac{M^2}{r^2}\frac{1}{1-2M/r}. \tag{2.3}$$

Assuming, for massless fields, that the only contribution to the trace of the energy-momentum tensor $T_\alpha^\alpha$ is the conformal anomaly, the trace is found directly from the Weyl tensor $C_{\alpha\beta\gamma\delta}$ and the Ricci tensor $R_{\alpha\beta}$ of the spacetime geometry,

$$T_\alpha^\alpha = \sum_s q_s \frac{m_P^2}{2880\pi^2}\left( C_{\alpha\beta\gamma\delta}C^{\alpha\beta\gamma\delta} + R_{\alpha\beta}R^{\alpha\beta} - \frac{1}{3}R^2 \right), \tag{2.4}$$

regardless of the quantum state of the fields. The coefficients are $q_0 = 1$ (single massless scalar field), $q_1 = -13$ (spin 1), and $q_2 = 212$ (spin 2)[26]. In the Schwarzschild geometry $R_{\alpha\beta} = R = 0$ and

$$C_{\alpha\beta\gamma\delta}C^{\alpha\beta\gamma\delta} = 48\frac{M^2}{r^6}, \tag{2.5}$$

so

$$T_\alpha^\alpha = -E + P_r + 2P_t = \sum_s q_s \frac{m_P^2}{60\pi^2}\frac{M^2}{r^6} = 64\sum_s q_s \sigma T_H^4 \left(\frac{2M}{r}\right)^6. \tag{2.6}$$

In considering how to interpret physically the energy-momentum tensor for a spin $s$ field, the obvious first step is to split off a "conformal anomaly" part from a "regular" part. These have quite different physical origins, and for the higher spins the conformal anomaly part strongly dominates in the vicinity of the horizon. The conformal anomaly part should be separately conserved and respect all of the symmetries of the Schwarzschild geometry, including the invariance of the Schwarzschild curvature tensor under radial boosts, which requires that $P_r^{ca} = -E^{ca}$, so $2P_t^{ca} = T_\alpha^\alpha - 2P_r^{ca}$. Equations (2.2) and (2.6) then give

$$E^{ca} = -P_r^{ca} = +\frac{1}{2}P_t^{ca} = \frac{1}{2}T_\alpha^\alpha = 32\sigma T_H^4 x^6 \sum_s q_s, \qquad (2.7)$$

$x \equiv 2M/r$, with the constant of integration chosen to eliminate a traceless contribution proportional to $x^4$. Christensen and Fulling[3] chose as a conformal anomaly part of the energy-momentum tensor one that is not invariant under radial boosts.

The traceless regular part of the energy-momentum tensor can be further decomposed into subparts. While Christensen and Fulling demanded that each part of their decomposition satisfy energy-momentum conservation, there is no real physical justification for doing so, particularly if one wants to understand how the Hawking radiation is generated. After all, generation implies a transfer of energy and momentum. What seems more appropriate in the Unruh vacuum case is to consider the regular part to be made up of an outgoing null fluid, with $E^{out} = F^{out} = P_r^{out}$, an ingoing null fluid with $E^{in} = -F^{in} = P_r^{in}$, and a "vacuum polarization" part with $E^{vp} = -P_r^{vp} = P_t^{vp}$. Any traceless spherically symmetric energy-momentum tensor can be split up in this way, but it is a good match to the physics to the extent that the flow of energy is actually predominantly radial. This form does not make physical sense for the Hartle-Hawking vacuum, for which an alternative decomposition is proposed below.

There is one free function in this decomposition not determined by Eqs. (2.1) and (2.2). I find it convenient to take this free function $f(x)$ as describing how the net energy flux of Eq. (2.3) is split into an outgoing part and an ingoing part such that $F^{out} = (1-f)F$ and $F^{in} = fF$. (Note that a positive ingoing flux requires a negative ingoing energy density.) In order that the energy-momentum tensor be smooth on the horizon as measured by a freely falling observer, $1-f(x)$ must approach zero quadratically as $x \to 1$. One factor of $1-x$ is needed to counter the same factor in the denominator of $F$. The second factor is because a finite outward flux in a freely falling frame becomes infinitely redshifted in the boost to the static frame as $r \to 2M$. While $F^{in}$ is then singular as $x \to 1$, this is because of the infinite blueshift in transforming an ingoing null fluid from the freely falling frame to to the static frame. To account for the restriction on $f$, I introduce a function $h_s(x)$ in terms of which, for a field of spin $s$,

$$f(x) = x^3 \left[ 4 - 3x - 4(1-x)^2 h_s(x) \right],$$
$$1 - f = (1-x)^2 \left( 1 + 2x + 3x^2 + 4x^3 h_s \right). \qquad (2.8)$$

Given $h_s(x)$, Eq. (2.2) determines $E^{vp}(x)$ within a constant of integration $D_s$,

$$E^{vp} = \sum_s 2k_s \sigma T_H^4 x^4 \left[ x(1-x)h_s + D_s - \int_0^x \left( 3 + (3x'-2)h_s \right) dx' \right]. \qquad (2.9)$$

Smooth behavior as $x \to 0$ requires that $f \sim x^3$ in this limit. A simple choice for $h_x$ is just a constant value

$$h_s(x) = c_s. \tag{2.10}$$

The integration constant $D_s$ is related to the angular spread of the Hawking radiation at large $r$, i.e., to the apparent angular size of the black hole as seen by a distant observer viewing the Hawking radiation.

Visser[10] made a careful analytic fit to the numerical results of Jensen, et al[9] in the case of a massless scalar field, based on a polynomial model for the transverse stress and their actual numerical data, rather than the published graphs. When translated into my formalism his analytic model, which agrees with the numerical results within their accuracy, about 1%, corresponds to $c_0 = 0.540$ and $D_0 = 0.621$. Without access to the numerical data, I can make only a crude fit to the graphs of the spin 1 results in Ref. [9]. One problem is that most of the spin 1 graphs are strongly dominated by the conformal anomaly contribution, and are very insensitive to the value of $D_1$. What seems to give a reasonable fit, accurate to about 10%, is $c_1 = 3.8$. A reasonable guess is that the value of $D_s$ doesn't change much with spin, since in all cases one expects the radius of the black hole image to be something like the impact parameter of a marginally trapped null geodesic trajectory, $3\sqrt{3}M$. I adopt the spin 0 value for all $D_s$ in presenting results. Finally, with no spin 2 numerical results in the literature, I guess that $c_2 = 25$ might be in the right ballpark, assuming that $c_s$ increases by about the same factor for each unit increase in spin.

Inspection of Eq. (2.8) and the plots of $f(x)$ in Fig. 1 show that $c_s > 1$ implies a negative $f$ at smaller $x$ (larger $r$). Where $f$ is negative the incoming component of the radiation has a net *positive* energy density. However, there is always a macroscopic region around the horizon where the energy flow in the static frame is dominated by inward flow of *negative* energy, which can only be understood as associated with macroscopic quantum behavior. As emphasized by Visser[10], all of the standard classical energy conditions used to prove the existence of event horizons and the existence of singularities inside event horizons are violated by the semi-classical energy-momentum tensor. As shown in Fig. 1, not too far inside the horizon the ingoing part of the local energy density becomes positive $(f < 0)$. However, analytic extrapolation inside the horizon of approximate formulas validated only outside the horizon is very questionable. According to the usual pair creation story, the energy flux well inside the horizon due to the Hawking partner quanta should have locally positive energy density propagating outward with respect to local freely falling observers along null geodesics with negative energy parameters with respect to the Schwarzschild "time" Killing vector, but it is not clear that a classical particle description of the Hawking partners is valid there, since the relevant modes have wavelengths comparable to the radius in the interior.

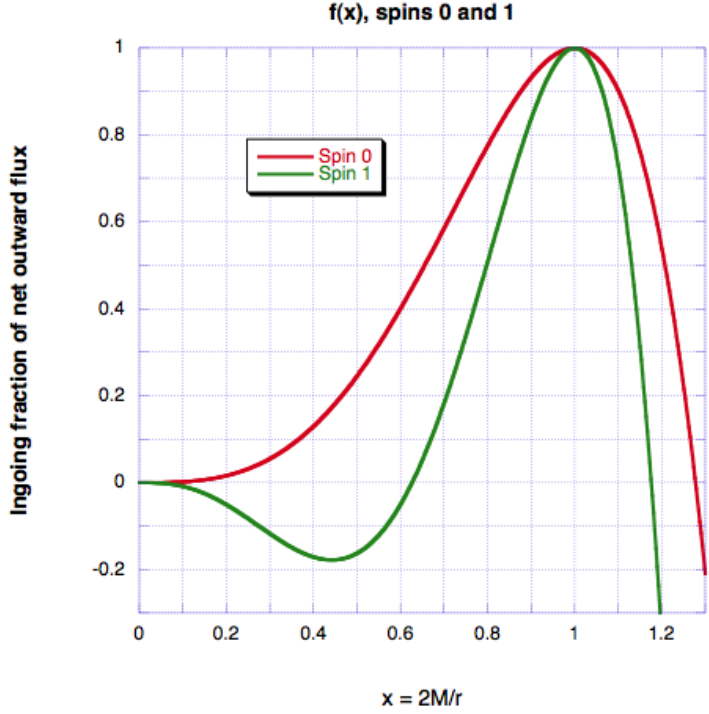

Figure 1. Plots of $f(x)$, the fraction of the net energy flux that is ingoing, for spin 0 and spin 1 fields. Positive values imply an ingoing component with negative energy density. The curves for $x > 1$ are questionable analytic extrapolations.

While the static frame ingoing energy density is infinitely negative infinitesimally close to $r = 2M$ and the static frame outgoing energy density is zero there, both are well behaved in a freely falling frame. Of course, there any many freely falling frames, but a convenient choice is the frame carried by an observer freely falling from rest at $r = \infty$. The 3-velocity of this freely falling observer in the local static frame is $v_{ff} = -\sqrt{2M/r} = -\sqrt{x}$. The Lorentz transformation to the free-fall frame gives

$$E_{ff}^{out} = \frac{1}{4} k_s \sigma T_H^4 \left(1 + \sqrt{x}\right)^2 x^2 \left(1 + 2x + 3x^2 + 4c_s x^3\right),$$
$$E_{ff}^{in} = -\frac{1}{4} k_s \sigma T_H^4 \left(1 + \sqrt{x}\right)^{-2} x^5 \left[4 - 3x - 4(1-x)^2 c_s\right].$$
(2.11)

The vacuum polarization part is unaffected by the change of frame,

$$E^{vp} = 2 k_s \sigma T_H^4 x^4 \left[D_s + 3(c_s - 1)x - \frac{5}{2} c_s x^2\right].$$
(2.12)

With this choice of free-fall frame, the net energy density is clearly positive for all $x$ in the range 0 to 1. The values of the different contributions to the energy density at the horizon for spins 0, 1, and 2 are compared in Table I. The fact that $E_{ff}^{out}$ is non-zero on the horizon does not mean any significant amount of Hawking radiation is generated there, since radiation is infinitely redshifted in propagating outward away from the horizon and the energy-momentum tensor there is dominated by quantum effects.

| Spin | $k_s$ | $q_s$ | $c_s$ | $E_{\text{ff}}^{\text{out}}$ | $E_{\text{ff}}^{\text{in}}$ | $E^{\text{vp}}$ | $E^{\text{ca}}$ |
|---|---|---|---|---|---|---|---|
| 0 | 14.26 | 1 | 0.54 | 116.4 | -0.891 | -60.15 | 32 |
| 1 | 6.49 | -13 | 3.8 | 137.6 | -0.406 | -6.22 | -416 |
| 2 | 0.742 | 212 | 25 | 79.7 | -0.046 | 15.0 | 6784 |

Table I. Coefficients in the analytic expressions and the corresponding energy densities at the horizon, $x = 1$, in the fiducial freely-falling frame. The energy densities are given in units of $\sigma T_{\text{H}}^4$.

The non-zero-spin energy densities are enhanced by a factor of two because massless spinning particles have two helicity states. The numbers in Table I should not be over-interpreted. The magnitudes of $E_{\text{ff}}^{\text{out}}$ and $E_{\text{ff}}^{\text{in}}$ relative to each other and relative to $E^{\text{vp}}$ are highly sensitive to the particular choice of a timelike free-fall observer. There is no indication of anything special happening right at the event horizon, and as Unruh has argued[5], the Hawking radiation should be considered as originating from the evolution of the vacuum modes of the *fields* rather than as a particle tunneling process. Wave packets behaving like particles can only be constructed well away from the horizon. What is significant is the large contribution of the spin 2 $E^{\text{ca}}$ to the total energy density at the horizon. While for a wide range of observers the net energy density, summed over spins, is positive everywhere, there are observers for which the net energy density is negative over some finite region around the horizon.

If the Parikh and Wilczek model[12] for the generation of the Hawking radiation as quantum tunneling across the horizon over a Planck scale distance were correct, it would seem a reasonable rough approximation to propagate the radiation over larger distances from the horizon as a classical null fluid, with an energy-momentum tensor of the form

$$T^{\alpha\beta} = \mu^{\text{out}} k^\alpha k^\beta, \tag{2.13}$$

$k^\alpha = dx^\alpha / d\lambda$ the tangent vector to an outgoing radial null geodesic with affine parameter $\lambda$. To calculate the evolution of this energy-momentum tensor along an outward radial null geodesic, I work in advanced Eddington-Finkelstein coordinates, since these are regular on the future horizon. With an advanced time coordinate

$$v = t + \int^r \frac{dr'}{1 - 2M/r'}$$

constant on ingoing radial null geodesics, the Schwarzschild metric becomes

$$ds^2 = -(1 - 2M/r)dv^2 + 2dvdr + r^2(d\theta^2 + \sin^2\theta \, d\varphi^2). \tag{2.14}$$

The outgoing radial null geodesic has

$$k^r = \frac{1}{2}\left(1 - \frac{2M}{r}\right)k^v, \quad k_v = -\frac{1}{2}\left(1 - \frac{2M}{r}\right)k^v, \quad k_r = k^v = \frac{dv}{d\lambda}. \tag{2.15}$$

Assuming free streaming of the null fluid, with no interactions,

$$T^\beta_{\alpha;\beta} = 0 = \frac{1}{r^2}\frac{d}{d\lambda}\left(r^2 \mu^{\text{out}}\right). \tag{2.16}$$

Since the metric is independent of $v$, $k_v$ is constant along the geodesic, and

$$\left(1-\frac{2M}{r}\right)\frac{dv}{d\lambda} = (1-x)\frac{dv}{d\lambda} = \text{constant} = e. \tag{2.17}$$

The energy-momentum coordinate components are

$$T_v^v = -\frac{1}{2}\frac{e^2\mu^{\text{out}}}{1-x}, \quad T_v^r = -\frac{1}{4}e^2\mu^{\text{out}}, \quad T_r^v = \frac{e^2\mu^{\text{out}}}{(1-x)^2}, \quad T_r^r = \frac{1}{2}\frac{e^2\mu^{\text{out}}}{1-x}. \tag{2.18}$$

The projection into the frame of the fiducial free-fall observer with 4-velocity in these coordinates $u^v = \left(1+\sqrt{x}\right)^{-1}$, $u^r = -\sqrt{x}$, gives

$$E_{\text{ff}}^{\text{out}} = \mu^{\text{out}}\left(u^\alpha k_\alpha\right)^2 = \frac{1}{4}\frac{e^2\mu^{\text{out}}}{\left(1-\sqrt{x}\right)^2}. \tag{2.19}$$

Since $E_{\text{ff}}^{\text{out}} \sim m_{\text{P}}^2/(M^2 r^2)$ well outside the horizon, extrapolation back to $r - 2M \sim m_{\text{p}}$ gives $E_{\text{ff}}^{\text{out}} \sim 1/M^2$ there. While this is not a super-Planckian firewall, it is a drastic difference from the semi-classical prediction and its interaction with any accreting matter would be easily observable.

A more physical decomposition of the Hartle-Hawking energy-momentum tensor might also be useful. Again, I work with physical components in the static frame. First, separate out a conformal anomaly part identical to that given for the Unruh vacuum in Eq. (2.7). Of course, the net radial energy flux $F$ vanishes. A natural decomposition of the traceless regular part is to start with a "thermal radiation" part, with an isotropic pressure $P^{\text{th}}$ and an energy density $E^{\text{th}} = 3P^{\text{th}}$. An explicit expression for $P^{\text{th}}$ based on the Page[7] analytic expressions for the scalar field energy-momentum tensor is

$$\begin{aligned}P^{\text{th}} &= \frac{2}{3}\sigma T_{\text{H}}^4 \left(1 + 2x + 3x^2 + 4x^3 + 5x^4 + 6x^5 - 21x^6\right) \\ &= \frac{2}{3}\sigma T_{\text{H}}^4 \frac{1 - x^6\left[7 - 6x + 21(1-x)^2\right]}{(1-x)^2}.\end{aligned} \tag{2.20}$$

This differs from the corresponding Page analytic formula, which has a coefficient 9 in place of the 21. The advantage of my definition is that it is regular on the horizon for a freely falling observer, since $P^{\text{th}} = 0$ at $x = 1$. It might be desirable to allow an additional contribution from radially counter-streaming null fluids with equal energy density $E^{\text{csn}}$ and radial stress $P_r^{\text{csn}}$ in the static frame, but zero net energy flux and transverse stress. In order that this piece be regular on the horizon in a freely-falling frame, I require that $E^{\text{csn}} = 0$ at $x = 1$. It is otherwise unrestricted, and in terms of an arbitrary bounded function $g(x)$

$$E^{\text{csn}} = \frac{2}{3}\sigma T_{\text{H}}^4 x^5 (1-x) g. \tag{2.21}$$

The Page analytic formula corresponds to $g = 0$, but a simple extension of the Page formula, perhaps allowing a better fit to numerical results, is to take $g$ to be a non-zero constant.

The remaining piece has the "vacuum polarization" form familiar from the Unruh vacuum decomposition. This is determined within a constant of integration from the momentum conservation equation, with the result

$$E^{vp} = \frac{2}{3}\sigma T_H^4 x^4 \left[ x(1-x)g + D + \int_0^x (2-3x')g\,dx' - 84x^2 \right]. \quad (2.22)$$

The constant $D$ is a free integration constant, set equal to zero in the Page formula. Note that $T_\theta^\theta = P_t = P^{th} + E^{vp} + 2E^{ca}$.

While the Page formula works reasonably well for zero spin, for higher spins both $g$ and $D$ will probably be required to get decent fits to numerical data. Also, modifications of the coefficients in the $P^{th}$ expression may be necessary for good results. See Jensen and Ottewill (1989) for a discussion of analytic formulae in the spin 1 case. The overall normalization in Eqs. (2.20), (2.21), and (2.22) should be increased by a factor of two for non-zero spins in order to account for their two helicity states.

## III. SEMI-CLASSICAL BACK-REACTION ON THE GEOMETRY

The expectation value of the renormalized quantum energy-momentum tensor can be inserted as a source in the classical Einstein equations to calculate first-order corrections to the classical spacetime geometry on which the calculation of the energy-momentum tensor was based. This procedure is not always justified. One can imagine a quantum event acting as a trigger for some large macroscopic rearrangement of matter that alters the gravitational field in a non-perturbatively. The outcome of the quantum event should in this case be understood as generating alternative classical histories, rather than perturbing a single classical history. But in the case of black hole evaporation it seems reasonable to calculate *first-order* corrections to the geometry of spacetime in this way. The extension of the semi-classical corrections to higher order in an expansion in powers of $\hbar$, while continuing to use an expectation value of the energy-momentum tensor in the classical Einstein equations, is a much more dubious proposition and in some sense is certainly false. As pointed out by Nomura, et al[27], the unitary evolution of a black hole must be considered at some point the superposition of many classical histories, due, for instance, to random recoil of the black hole from emission of Hawking particles in random directions.

With these caveats, I now consider the corrections to the geometry first order in $\hbar$, both outside and inside the Schwarzschild horizon, working in advanced Eddington-Finkelstein coordinates $(v,r)$ with $v$ constant along ingoing radial null geodesics. A general spherically symmetric form of the metric in these coordinates is

$$ds^2 = -Ae^{2\psi}dv^2 + 2e^\psi dvdr + r^2\left(d\theta^2 + \sin^2\theta\,d\varphi^2\right), \quad (3.1)$$

following the notation of Bardeen[13]. The inverse metric has

$$g^{vv} = 0, \quad g^{vr} = e^{-\psi}, \quad g^{rr} = A \equiv 1 - \frac{2m}{r}. \quad (3.2)$$

The Einstein equations can be put in the simple form

$$\left(\frac{\partial m}{\partial v}\right)_r = 4\pi r^2 T_v^r, \quad \left(\frac{\partial m}{\partial r}\right)_v = -4\pi r^2 T_v^v, \quad \left(\frac{\partial \psi}{\partial r}\right)_v = 4\pi r e^\psi T_r^v. \tag{3.3}$$

The more complicated $G_\theta^\theta = 8\pi T_\theta^\theta$ equation is redundant, and $T_r^r \equiv T_v^v + Ae^\psi T_r^v$.

Before computing the back-reaction, we need the $v,r$ coordinate components of the energy-momentum tensor on the Schwarzschild background, where $\psi = 0$ and $A = 1 - 2M/r$. Outside the horizon, in terms of the static frame physical components from in Section II,

$$T_v^r = -(1-x)F = -\frac{1}{4}\sigma T_H^4 x^2 \sum_s k_s, \tag{3.4}$$

$$T_v^v = -E - F = -2E^{\text{out}} - E^{\text{vp}} - E^{\text{ca}}, \tag{3.5}$$

$$(1-x)T_r^v = E + P_r + 2F = 4E^{\text{out}}, \tag{3.6}$$

$$T_r^r = F + P_r = 2E^{\text{out}} - E^{\text{vp}} - E^{\text{ca}}, \tag{3.7}$$

$$T_\theta^\theta = P_t = E^{\text{vp}} + 2E^{\text{ca}}. \tag{3.8}$$

In particular,

$$T_r^v = \sigma T_H^4 x^2 \sum_s k_s \left(1 + 2x + 3x^2 + 4c_s x^3\right) \tag{3.9}$$

and

$$-T_v^v = \frac{1}{2}\sigma T_H^4 x^2 \sum_s k_s \left[1 + x + (4D_s + 1)x^2 + (16c_s - 15)x^3 - 14c_s x^4\right] + E^{\text{ca}}. \tag{3.10}$$

Only $E^{\text{ca}}$ as given by Eq. (2.7) can be extrapolated with confidence inside the horizon, since it is invariant under radial boosts and depends only on the background curvature. Analytically continuing the coordinate components of the rest of the energy-momentum tensor to $x > 1$, with the possible exception of (3.4), does not make physical sense. The natural analogue of the exterior static frame inside the horizon is the "Kasner" frame of an observer whose four-velocity is orthogonal to a constant-$r$ hypersurface. The relationship between the coordinate components and the Kasner-frame physical components is identical to that of Eqs. (3.4)-(3.8), except for the signs of $E^{\text{out}}$ in Eqs. (3.5) and (3.7) and using $|1-x|$ in Eqs. (3.4) and (3.6).

The key point is that $\partial m / \partial v$ is independent of $r$ outside the horizon, and if extrapolation of $T_v^r$ is valid, inside as well, suggesting that the cumulative effect of the Hawking radiation is to gradually decrease the Schwarzschild mass, while preserving the Schwarzschild geometry except for corrections at most of order $(m_p / M)^2 \log(M/m_p)$. The logarithm is due to a logarithmic divergence of the integral for $\psi$, where on the past segment of the incoming null hypersurface Hawking radiation is present, assuming that $\psi = 0$ at past null infinity.

The upshot, as argued by Bardeen[13], is that there is no indication within the standard semi-classical results of any breakdown of that approximation in the vicinity of or outside of the horizon as the black hole evaporates as long as the black hole mass satisfies $M/m_p \gg 1$. The quantum fields should continue to fluctuate near the horizon as expected for the Unruh vacuum in a Schwarzschild background, and the semi-classical energy-momentum tensor should remain valid. There are issues with the fact that the modes giving rise to the Hawking radiation after even a small amount of evaporation had sub-Planckian wavelengths when extrapolated back to past null infinity, but similar issues arise in inflation in the early universe without any apparent effect on the spectrum of the CMB fluctuations.

What about the deep interior of the black hole? Since I don't believe quantum information can be stored in a stretched horizon, if the quantum information in the collapsing star or accumulating with the Hawking partners is absorbed by a singularity I see no way to argue for preservation of unitarity unless one is willing to give up the principle of locality in quantum field theory. Can quantum back-reaction drastically alter the geometry deep inside the horizon and prevent formation of a singularity? The correct effective energy-momentum tensor deep inside the horizon is not known, except for the conformal anomaly piece. However, when gravitons are included, the conformal anomaly strongly dominates the rest of the energy-momentum tensor in the vicinity of the horizon, and quite possibly would do so even in the deep interior. Therefore, it is interesting to look at the back-reaction from just the conformal anomaly part of the effective energy-momentum tensor.

Let $q^{tot} = \sum_{s=1,2} q_s = 199$. Then from Eqs. (2.7), (3.5), and (3.3),

$$m = M - \frac{4\pi}{3} 32 q^{tot} \sigma T_H^4 \frac{(2M)^6}{r^3} = M - \frac{q^{tot}}{90\pi} \frac{m_p^2 M^2}{r^3}. \tag{3.11}$$

The mass function is substantially modified once $r \lesssim (m_p^2 M)^{1/3}$, which, while much larger than the Planck length for a large black hole, is where the curvature becomes Planckian. Any extrapolation beyond this point cannot rely on semi-classical theory.

A particular non-singular model for the interior of an asymptotically Schwarzschild black hole that has been explored by Hayward[28] and Frolov[29],

$$m = \frac{M(v)r^3}{r^3 + 2a^2 M(v)}, \tag{3.12}$$

with $a$ a constant the order of the Planck length. This matches Eq. (3.11) for $a^2 = q^{tot} m_p^2 /(180\pi)$. The geometry is Schwarzschild-like with mass $M$ for $r \gg (a^2 M)^{1/3}$ and de Sitter-like for $r \ll (a^2 M)^{1/3}$. An inner apparent horizon is located at $r \cong a$ when $M \gg m_p$. Taking $M(v)$ to decrease in accord with the Hawking luminosity formula eventually results in the removal of all trapped surfaces (no zeroes of $1 - 2m/r$), but not until $M \sim a \sim m_p$. This is not satisfactory for resolving the information paradox, since just before the trapped surfaces disappear the black hole would have a von Neumann entropy vastly exceeding its Bekenstein-Hawking entropy.

A form similar to Eq. (3.12) has been suggested by Bonanno and Reuter[30], based on a renormalization group approach to quantum gravity:

$$m = \frac{Mr^3}{r^3 + \tilde{\omega} m_p^2 (r + \gamma M)}, \tag{3.13}$$

with $\tilde{\omega}$ and $\gamma$ constants of order one. As long as $\gamma$ is not zero, there is no qualitative difference from Eq. (3.12) in the deep interior of the black hole.

## IV. A SCENARIO FOR TRAPPED SURFACES DISAPPEARING BEFORE THE PAGE TIME

The most difficult challenge for singularity avoidance in the formation and evolution of a black hole is in the interior of the collapsing star, since the positive stresses of the matter energy-momentum tensor would offset possible negative stresses associated with quantum back-reaction. A number of authors have argued on various grounds that a bounce will occur around the Planck scale. There are various ideas about the implications of such a bounce for the causal structure of the spacetime and the fate of an evaporating black hole (see, e.g., Ref. [30], Ashtekar and Bojowald[31], Hossenfelder, et al[32], and Torres and Fayos[33]). What I would hope to be the result is that the entire star is swept up into a non-singular core to the black hole, of initially very small, but substantially larger than Planckian, size, in which there are no trapped surfaces. Then the subsequent evolution might be described as the evolution of a quasi-classical spacetime geometry, albeit with very large quantum back-reaction, justified by a macroscopic averaging over the large number of quanta in the core. A reasonable definition of the outer boundary of the core is the inner boundary of the trapped surfaces in the geometry, i.e., the inner apparent horizon. Matter and radiation (including the Hawking partners) can go inward across the inner apparent horizon, but once inside must remain inside as long as the inner apparent horizon exists. In constructing a model scenario, I make the drastic simplifying assumption that spherical symmetry continues to hold, even though small deviations from spherical symmetry are greatly amplified in gravitational collapse.

Without a singularity at $r = 0$ it is possible to appeal to Bousso's covariant entropy bound conjecture[23] to say something about how this core might evolve. The covariant entropy bound applies to a two-surface $B$. There are four families of null geodesics orthogonal to the two-surface, "inward" future-directed and past-directed null geodesics and "outward" future-directed and past-directed null geodesics. Each family forms a null hypersurface, or "light-sheet", which ends in caustics. The covariant entropy bound states that for any such null hypersurface whose expansion away from $B$ is negative, i.e., the area of two-surfaces of constant affine parameter from $B$ decreases away from $B$, the number of quantum degrees of freedom (the von Neumann entropy $S_{vN}$) crossing the null hyperface should satisfy the bound $S_{vN} \leq A/4$ (in Planck units), where $A$ is the area of the two-surface $B$. While various proofs of the conjecture have been attempted (e.g., Flanagan et al[34] and Bousso et al[35]), none of them apply in the presence of large quantum back-reaction, as is the case here, and entropy and energy are inherently non-local concepts in quantum field theory. Aside from these fundamental limitations of the Bousso bound, one can certainly object that large deviations from spherical symmetry are likely present in the core for a generic black hole, leading to caustics preventing any extension of the light-sheet to close to $r = 0$. Still, I would argue that the Bousso bound applies at least in spirit to a light–sheet generated by the inward radial null geodesics orthogonal to a sphere at or just outside the inner apparent horizon.

Bousso's original discussion of the covariant entropy bound in the context of gravitational collapse was based on the conventional picture that a collapsing star forming a black hole is crushed to an infinite density singularity on a spacelike hypersurface. The ingoing light-sheet from a closed two-surface inside or outside the star near $r = 0$ will only intersect at most a small portion of the star before it terminates at the singularity, even if it does not terminate sooner as the result of caustics generated by inhomogeneities. So while the area in Plank units of the two-surface may be much less than the von Neumann entropy of the entire star, the von Neumann entropy intersected by the light-sheet would be consistent with the covariant bound. However, in the scenario I am postulating there is no singularity, and all of the von Neumann entropy of the star ends up inside the core, i.e., inside the inner apparent horizon. The light-sheet at constant advanced time $v$ inside the inner apparent horizon intersects essentially all of the von Neumann entropy in the core. While small deviations from spherical symmetry will cause caustics before the light sheet reaches all the way to $r = 0$, most of the entropy should be concentrated close to the inner apparent horizon, based on the behavior of null geodesics in the core. The von Neumann entropy of the core is initially tiny compared to the Bekenstein-Hawking entropy associated with the outer apparent horizon, but as long as it is non-zero the covariant bound implies a minimum to the area of the inner apparent horizon.

As the number of quantum degrees of freedom in the core increases due to the core trapping the Hawking partners being created as the black hole evaporates, the radius of the core must increase. When the von Neumann entropy of the trapped quanta gets close to the Bekenstein-Hawking entropy associated with the outer apparent horizon of the black hole, how close depending on how close the Bousso bound is to being saturated, the inner and outer horizons merge, which means the trapped surfaces have disappeared. This is at or before the Page time. Note that if the system started out in a pure state, the von Neumann entropy of the core is also an entanglement entropy. Depending on whether the star is in itself a pure or state or entangled with qbits in the external universe, the initial von Neumann entropy of the core may or may not be zero.

In the context of the Hayward form of $m(v,r)$ (Eq. (3.12)), the inner apparent horizon is at $r \cong a$ as long as $a \ll M$. What is required for my scenario is that, instead of being a constant, $a$ should increase with advanced time as the Hawking partners accumulate in the core, in order to maintain $\pi a^2$ greater than or equal to the von Neumann entropy of the core, $S_{vN}$. If $a^2$ is a function only of advanced time $v$, the effective energy-momentum tensor in the vicinity of the outer apparent horizon will reflect the same value of $a^2$, and differ from what is required to fit the semi-classical conformal anomaly. This is not unreasonable if the effect is due to the entanglement between the Hawking radiation quanta and their partners causing an enhanced quantum gravitational contribution to the back-reaction. The basic semi-classical framework could remain in place in the vicinity of the outer apparent horizon as long as $a \ll M$, except for the *assumption* of time independence made in deriving the energy-momentum tensor. The entanglement grows as the Hawking pairs are created, and strongly affects the core as the Hawking partners accumulate there.

It is not probably not correct to model the Hawking partners as having locally negative energy propagating along inward radial null geodesics, as implied by the Hayward metric with $M = M(v)$ and $a^2 = a^2(v)$. If the Hawking partners have locally positive energy, they must propagate toward the outward edge of the light cone, implying a non-zero metric function $\psi$. They still reach the core, but not as quickly, as discussed at the end of this section.

For my extension of the Hayward model, the time dependence of $M$ should still correspond to the decrease in black hole mass given by the Hawking-Page formula of Eq. (1.4). The von Neumann entropy $S_{vN}$ accumulating in the core of the black hole is dominated by the Hawking partners, and their von Neumann entropy at advanced time $v$ should correspond to the von Neumann entropy of the Hawking radiation outside the black hole calculated roughly at the *retarded* time the constant-$v$ hypersurface reaches $r \approx 3M$. Page's estimate[14] of the latter entropy, assuming the radiation von Neumann entropy is roughly equal to its thermodynamic entropy, gives

$$\frac{dS_{vN}}{dv} \approx \frac{1}{715M}. \tag{4.1}$$

Assuming the Bousso covariant entropy bound applied to the inner apparent horizon is close to being saturated, we want $\pi a^2 \approx m_P^2 S_{vN}$ and therefore

$$\frac{da^2}{dv} \approx \frac{1}{2246}\frac{m_P^2}{M} \equiv 2\pi\beta(2M)^3 \sigma T_H^4, \tag{4.2}$$

$\beta = 21.5$.

From the $m(v,r)$ of Eq. (3.12), and with $\sum_s k_s \equiv \alpha$, substitution into the Einstein equations gives

$$-T_v^v = \frac{1}{4\pi r^2}\frac{\partial m}{\partial r} = \frac{3}{2\pi}\frac{a^2 M^2}{\left[r^3 + 2a^2 M\right]^2}, \tag{4.3}$$

$$-T_v^r = -\frac{1}{4\pi r^2}\frac{\partial m}{\partial v} = \sigma T_H^4 M^2 \frac{\alpha r^4 + \beta(2M)^3 r}{\left[r^3 + 2a^2 M\right]^2}, \tag{4.4}$$

$$T_r^r = T_v^v, \quad T_r^v = 0. \tag{4.5}$$

The general spherically symmetric momentum conservation equation $T^\alpha_{r;\alpha} = 0$ is

$$T_{r,v}^v + \frac{e^\psi}{r^2}\left(r^2 e^{-\psi} T_r^r\right)_{,r} + \left(\frac{1}{2}A_{,r} + \psi_{,r}\right)T_r^v - \frac{2}{r}T_\theta^\theta = 0, \tag{4.6}$$

which for the generalized Hayward model gives

$$T_\theta^\theta = T_r^r + \frac{r}{2}\frac{\partial T_r^r}{\partial r}. \tag{4.7}$$

This energy-momentum tensor of Eqs. (4.3)-(4.5) at best only applies in the *interior* of an evaporating black hole. The geometry transitions from de Sitter-like for $r \ll (2a^2 M)^{1/3}$ to Schwarzschild-like for $r \gg (2a^2 M)^{1/3}$ as long as $a \ll M$.

The interior energy-momentum tensor should transition to an exterior energy-momentum tensor that at large $r$ has positive energy propagating outward rather than negative energy propagating inward. While Hayward joined the advanced Vaidya-like metric to a retarded Vaidya-like metric at a surface layer somewhat outside the horizon, as long as $a \ll M$ an exterior semi-classical energy-momentum tensor analogous to that of Section II can be constructed that transitions to the interior energy-momentum tensor at $r = 2M$ without a surface layer. The absence of a surface layer requires that $T_r^r$ and $T_v^r$ be continuous across $r = 2M$, but other components, such as $T_\theta^\theta$, can be discontinuous.

The first-order interior energy-momentum tensor is

$$T_v^v = T_r^r = -\frac{3}{2\pi} \frac{a^2 M^2}{r^6} = -5760 \sigma T_H^4 \left(\frac{\pi a^2}{m_P^2}\right) x^6, \quad T_r^v = 0,$$

$$-T_v^r = \sigma T_H^4 \frac{M^2}{r^2}(\alpha + \beta x^3) = \frac{1}{4} \sigma T_H^4 (\alpha x^2 + \beta x^5), \quad T_\theta^\theta = -2 T_r^r.$$

(4.8)

The radial dependence of the diagonal components has the form of the conformal anomaly discussed in Section II, but for $\pi a^2 \gg m_P^2$ the actual conformal anomaly is only a small constant contribution to $a^2$. Note that $\pi a^2 / m_P^2$ is just the von Neumann entropy of the core, assuming the Bousso bound is saturated. The first term in $T_v^r$ arises from $M(v)$, while the second is associated with the time dependence of $a^2$ in $T_v^v$ through the $T_{v;\beta}^\beta = 0$ conservation equation. The second term is about three times larger than the first at $x = 1$, and totally dominates in the deep interior.

The form of the interior energy-momentum tensor is consistent with the idea that what enforces the Bousso bound and the growth of the core might be the growing number of entangled Hawking-Hawking partner pairs. If each pair is connected by a microscopic Einstein-Rosen bridge, in the spirit of the ER=EPR conjecture of Maldacena and Susskind[22], it is not unreasonable that the microscopic ER bridges modify the macroscopic gravitational field. While the radial dependence of the diagonal components has the same form as the semi-classical conformal anomaly, their magnitude is much larger and depends on the quantum state. Such non-anomalous contributions to the trace of the energy-momentum tensor are not out of the question, since the classical theory of general relativity is not conformally invariant.

I now try to make an educated guess at a plausible form for the exterior energy-momentum tensor. It seems reasonable that the dominant contribution to $T_r^r$ should still have the form given in Eq. (4.8), but the parameters $M$ and $a^2$ should not just depend on the advanced time. Instead they should depend on a time $\tilde{t}$ that smoothly interpolates between the advanced time $v$ at $r = 2M$ to a retarded time at $r \gg 2M$, since at large $r$ the Hawking quanta propagate outward at the speed of light. Something like

$$\tilde{t} = v - 2 \int_{2M}^{r} w(r') dr', \quad w(r) = \frac{r(r-2M)}{(r-2M)^2 + \delta^2 (2M)^2}.$$

(4.9)

with $\delta = O(1)$ adjustable to give the proper match between the advanced time of the Hawking partner and the asymptotic retarded time of the Hawking radiation quantum with which it is entangled. To first-order in $(m_P / M)^2$, at $r$ not *very* much larger than $2M$, the time dependence of $M$ in Eq. (4.9) can be neglected. Then

$$\frac{\partial a^2}{\partial v} \cong \frac{da^2}{d\tilde{t}}, \quad \frac{\partial a^2}{\partial r} \cong -2w(r) \frac{da^2}{d\tilde{t}}.$$

(4.10)

In advanced Eddington-Finkelstein coordinates on the Schwarzschild background, the coordinate components $T_\alpha^\beta$ in the exterior can be written in terms of the same static frame energy densities invoked in Section II, with

$$-T_v^v = 2E^{\text{out}} + E^{\text{vp}} + E^{\text{ca}}, \quad T_r^r = 2E^{\text{out}} - E^{\text{vp}} - E^{\text{ca}},$$
$$T_v^r = (E^{\text{in}} - E^{\text{out}})(1-x), \quad T_r^v = 4E^{\text{out}}/(1-x), \quad T_\theta^\theta = E^{\text{vp}} + P_t^{\text{ca}}. \tag{4.11}$$

The strategy is to make ansatzes for $E^{\text{ca}}$ and $E^{\text{out}}$, with $E^{\text{ca}}$ including the postulated non-anomalous contribution to the trace of the energy-momentum tensor. Then $P_t^{\text{ca}}$ is determined from conservation of the "conformal anomaly" piece, and $E^{\text{vp}}$ is found from the $T_{r;\beta}^\beta = 0$ conservation equation for the rest of the energy-momentum tensor, which to first order is

$$\frac{x}{2}T_r^v - x^3 \frac{\partial}{\partial x}\left(\frac{1}{x^2}T_r^r\right) - 2T_\theta^\theta = 0. \tag{4.12}$$

To enforce the continuity requirements at $r = 2M$, I assume the expressions for $T_v^v$ and $E^{\text{ca}}$ as given by Eq. (4.8) for the interior, where $E^{\text{ca}} = -T_v^v$, also apply in the exterior, consistent with $T_{v;\beta}^\beta = 0$ to first order. I also require that $T_r^v$ be finite and $E^{\text{vp}}$ vanish at $x = 1$. In order that the energy flux be purely outgoing at large $r$, I want $E^{\text{out}} \cong -T_v^v$ as $x \to 0$.

The conservation of the "conformal anomaly" part gives

$$P_t^{\text{ca}} = \frac{1}{2}x^3 \frac{\partial}{\partial x}\left(\frac{1}{x^2}E^{\text{ca}}\right) = 2E^{\text{ca}} + \frac{3}{4}\beta\sigma T_H^4 x^5 w. \tag{4.13}$$

This is the only place where the distinction between $v$ and $\tilde{t}$ is significant at first order. I adopt a polynomial ansatz for $E^{\text{out}}(x)$. A form that allows the desired behavior at $x = 0$ and $x = 1$ and agrees with the semi-classical ansatz for the Hawking radiation in Section II is

$$E^{\text{out}} = \frac{1}{4}(1-x)\sigma T_H^4 x^2 \left\{\sum_s \left[k_s\left(1 + 2x + 3x^2 + 4c_s x^3\right)\right] + \beta x^3\left(1 + b_1 x + b_2 x^2\right)\right\}. \tag{4.14}$$

Substitute this into Eq. (4.12) to get an equation for $E^{\text{vp}}(x)$,

$$x^5 \frac{\partial}{\partial x}\left(\frac{E^{\text{vp}}}{x^4}\right) = \frac{2x^3}{1-x}\frac{\partial}{\partial x}\left[\frac{1}{x^2}(1-x)E^{\text{out}}\right]$$
$$= \sigma T_H^4 x^5 \left\{\begin{array}{l} \sum_s k_s\left[6(c_s - 1) - 10c_s x\right] + \\ \frac{1}{2}\beta\left[3 + 4b_1 x + 5b_2 x^2 - x\left(5 + 6b_1 x + 7b_2 x^2\right)\right] \end{array}\right\}. \tag{4.15}$$

The result is

$$E^{\text{vp}} = \sigma T_H^4 x^4 \left\{\begin{array}{l} 2\sum_s k_s\left[D_s + 3(c_s - 1)x - \frac{5}{2}c_s x^2\right] + \\ \frac{1}{2}\beta x\left[3 - \frac{5}{2}x + 2x(1-x)b_1 + x^2\left(\frac{5}{3} - \frac{7}{4}x\right)b_2\right] \end{array}\right\}. \tag{4.16}$$

The Hawking radiation piece is identical to that of Section II. None of the constant of integration is assigned to the entanglement piece, since I don't see why the entanglement contribution to $E^{vp}$ should fall off more slowly at large $r$ than the entanglement contribution to $T_v^r$. Continuity of $T_r^r$ at $x = 1$ requires that $E^{vp} = 0$ there. The spin 1 and spin 2 Hawking radiation pieces largely offset each other at $x = 1$, so the coefficient $b_2$ should be about 6 for consistency with the ansatz for the interior solution. The parameters $b_1$ here and $\delta$ in Eq. (4.9) could in principle be adjusted to optimize the fit an actual calculation of the entanglement energy-momentum tensor.

Exactly how the scenario might play out as $a$ becomes comparable to $M$ is beyond the scope of my semi-classical estimates. With the ansatz of Eq. (3.12), the inner and outer apparent horizons merge and disappear at $r = 1.33M$ when $a = 1.09M$. However, as the horizons merge, their surface gravity goes to zero, which naively could mean the rates of energy loss and increase in $a$ would both go to zero. Could the black hole persist as some sort of macroscopic remnant with a degenerate horizon? It seems perhaps more likely and certainly more desirable that instead the horizons should disappear completely, so all of the quantum information trapped inside the black hole can escape. The black hole would then be like a very long-lived resonance in a very inelastic scattering process accompanied by emission of huge numbers of soft photons and gravitons. The black hole might end in some sort of an explosion, but with energy densities comparable to the energy densities in the original collapsing star as the apparent horizon formed, not the super-Planckian energy density of the firewall scenario. The matter might emerge in a quite different form than in the original star. There is no reason to assume baryon number or lepton number would be preserved, for instance.

A schematic diagram of the early stages of the scenario is given in Figure 2, for the simplified case of a black hole formed by the collapse of a null shell, whose trajectory is the green line. Below the shell trajectory the spacetime is flat. The diagram has as a horizontal axis the radius r defined by the area of two-spheres. The diagonal lines are lines of constant advanced time $v$, and the advanced time at $r = 0$, for the mass function of Eq. (3.12) and $\psi = 0$, is the proper time of an observer there. The outer apparent horizon is plotted ignoring decrease in the Schwarzschild mass $M$, and the growth of the radius of the inner horizon formed as the collapsing shell reaches $r = 0$ is enormously exaggerated. Curving away from the outer apparent horizon are the trajectories of exterior and interior outward radial null geodesics, which in a crude way represent the trajectories of a Hawking particle and its partner. After the initial collapse the inner apparent horizon is a spacelike hypersurface, so the ingoing partner can cross the inner horizon into the core. By construction, the outer edge of the light cone is a vertical arrow on both apparent horizons.

A Penrose diagram for this scenario looks identical to the Minkowski Penrose diagram, since all null geodesics eventually end up at future null infinity. The large differences in radial propagation of null geodesics are hidden by the extreme conformal transformation necessary to make the radial null trajectories in the black hole Penrose diagram straight lines.

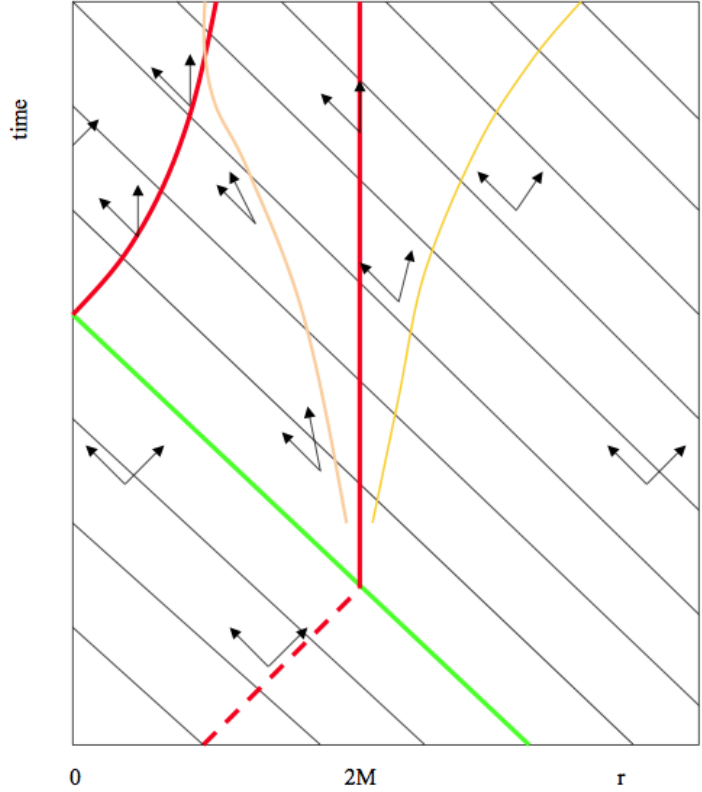

Figure 2. A schematic diagram of the early evolution of the black hole in my scenario. The diagonal lines are lines of constant advanced time $v$, the ingoing radial null geodesics, and the radius $r$ is constant on vertical lines. The green line represents the collapse of a shell to form the black hole. The right and left curved brown lines indicate trajectories of a Hawking quantum and its partner, respectively. The solid red lines are the two apparent horizons, with the radius of the inner horizon greatly exaggerated. The dashed red line is what would be the beginning of the event horizon, if one existed. The arrows indicate the light cones.

Unlike the outer apparent horizon, from which radial null geodesics diverge, radial null geodesics converge toward the inner apparent horizon to the extent that it is close to being a null hypersurface. There is a tendency for matter and radiation propagating near the outer edge of the light cone to pile up near the inner apparent horizon, but as long as there is a component of the energy flux propagating inward with negative energy density, as is true semi-classically at the outer apparent horizon and in the toy model is predominantly due to the time dependence of $a^2$, the the inner apparent horizon does not actually become singular.

It may seem a bit strange that the outward radial null geodesic can cross the inner apparent horizon, since in a time-independent geometry wth metric (3.1) the component $p_v$ of the geodesic tangent vector $p^\alpha$ is conserved and must be positive where there are trapped surfaces in order that $0 > dr/d\lambda = p^r = A p_r + e^{-\psi} p_v = -e^{-\psi} p_v$. But inside the inner apparent horizon all future-directed geodesics must have a positive energy parameter, or $p_v < 0$. The resolution is that the metric is actually time-dependent, so $p_v$ is not conserved and can change sign along with $A$ at the inner apparent horizon.

A careful look at the self-consistent propagation of radiation in the geometry as modified by the radiation sheds light on what is required. Suppose that the *outward* portion of the energy flux (presumably the Hawking partners) can be modeled as a flux of free-streaming classical radiation along the outward radial null direction, with an energy-momentum tensor $T_\alpha^{(\text{out})\beta} = \mu^{\text{out}} p_\alpha p^\beta$, where $p^\alpha$ satisfies the geodesic equation with the metric of Eq. (3.1). Then

$$p_{v;\beta}p^\beta = \frac{dp_v}{d\lambda} - \frac{1}{2}g_{\alpha\beta,v}p^\alpha p^\beta = \frac{dp_v}{d\lambda} + e^{2\psi}\left(\frac{1}{2}A_{,v} + A\psi_{,v}\right)(p^v)^2 - e^\psi \psi_{,v}p^v p^r = 0,$$

so

$$\frac{dp_v}{dv} = \frac{1}{p^v}\frac{dp_v}{d\lambda} = \frac{m_{,v}}{r}e^{2\psi}p^v - \psi_{,v}\left(Ae^{2\psi}p^v - e^\psi p^r\right). \tag{4.17}$$

Expressing all components of the tangent vector in terms of $p_v$, Eq. (4.17) becomes

$$\frac{dp_v}{dv} = -\frac{2p_v}{Ar}\frac{\partial m}{\partial v} + p_v\frac{\partial \psi}{\partial v} = -\frac{2p_v}{Ar}\frac{\partial m}{\partial v} - \frac{1}{2}Ae^\psi p_v\frac{\partial \psi}{\partial r} + p_v\frac{d\psi}{dr}. \tag{4.18}$$

The Einstein equations in the form of Eq. (3.3) give expressions for $m_{,v}$ and $\psi_{,r}$, from which

$$\frac{dp_v}{dv} - p_v\frac{d\psi}{dv} = -8\pi r\frac{p_v}{A}T_v^r - 2\pi r e^{2\psi}p_v A T_r^v. \tag{4.19}$$

Evaluating the energy-momentum tensor components,

$$T_v^r = T_v^{(\text{in})r} - \mu^{\text{out}}p_v^2 e^{-\psi}, \quad T_r^v = \mu^{\text{out}}\frac{4p_v^2}{A^2}e^{-3\psi}. \tag{4.20}$$

The "out" contribution cancels in Eq. (4.19), leaving

$$\frac{d}{dv}\left(e^{-\psi}p_v\right) = -8\pi r\frac{e^{-\psi}p_v}{A}T_v^{(\text{in})r}. \tag{4.21}$$

The Hawking partners initially have $p_v > 0$ somewhat inside the outer apparent horizon, where $A < 0$.

Now calculate how $A$ varies along the trajectory,

$$\frac{dA}{dv} = -\frac{2}{r}\left(\frac{\partial m}{\partial v} + \frac{A}{2}e^\psi\frac{\partial m}{\partial r}\right) + Ae^\psi\frac{m}{r^2}.$$

From the Einstein equations, and taking into account the vacuum polarization and "conformal anomaly" contributions to $T_v^v$ (denoted here just by $E^{\text{ca}}$), one sees that outgoing flux terms again cancel, and

$$\frac{dA}{dv} = 8\pi r T_v^{(\text{in})r} - 4\pi r Ae^\psi E^{\text{ca}} + Ae^\psi\frac{m}{r^2}. \tag{4.22}$$

The dominant contribution to $T_v^{(\text{in})r}$ in the trapped surface region is the term with coefficient $\beta$ in Eq. (4.4), of order $m_p^2/Ma^3$. Approaching the inner apparent horizon $E^{\text{ca}} \cong 3/(8\pi a^2)$ and $m \cong r^3/(2a^2)$. While $A \ll -1$ at the transition from the Schwarzschild-like to de Sitter-like, $|A|$ decreases exponentially with advanced time, going as $\exp(-ve^\psi/a)$ as $r \to a$ until $|A| < T_v^{(\text{in})r}/E^{\text{ca}} \sim m_\text{p}^2/(aM)$. From there the ratio $e^{-\psi}p_v/A$ stays approximately constant as both go through zero and change sign over an additional $e^\psi \Delta v \sim Ma^2/m_\text{p}^2$, which while many dynamical times is small compared with the evaporation time of $M^3/m_\text{p}^2$ as long as $a \ll M$. The metric function $\psi$ is estimated to be only of order $a/M$ during the transition.

## V. DISCUSSION

The scenario for black hole evaporation presented in this paper is rather speculative, but I believe it requires less drastic assumptions about new physics than alternative scenarios, whether or not they involve firewalls. It is quasi-classical in the sense that all stages of the evaporation can be described as quantum fluctuations about background classical spacetimes, though it is certainly not conventional semi-classical field theory with a fixed classical background. The effective quantum energy-momentum tensor is required to exert a very large back-reaction on the geometry over macroscopic distances, and in the proposed scenario this primarily occurs through a contribution to the energy-momentum tensor of the same form as, but much larger amplitude than, the semi-classical conformal anomaly. While the growth in back-reaction can be framed in terms of the Bousso covariant entropy bound, the growing amount of entanglement between the Hawking radiation quanta and the associated Hawking partners trapped by the black hole as the evaporation proceeds is invoked as the physical reason for the enhancement. The ER=EPR hypothesis of Maldacena and Susskind[22] suggests a modification to the quantum gravitational spacetime foam in the form of microscopic non-traversable Einstein-Rosen bridges connecting entangled particles. That such an effect occurs in certain specific AdS-CFT models has been verified by Jensen and Karch[36] and Jensen, et al[37], though the entanglement is between particles of the CFT on the AdS boundary, with only the Einstein-Rosen bridge in the bulk. While it seems plausible that, if it can be applied to mutually entangled photons and gravitons, this sort of structure should contribute to the macroscopic effective energy-momentum tensor in the interior of an evaporating black hole, I make no claim of any rigorous justification, even when the back-reaction on the geometry is small, so the macroscopic spacetime geometry can be approximated by the Schwarzschild metric. That quantum gravitational effects can extend over distances very large compared to the Planck scale seems well established in the semi-classical calculations of black hole evaporation, despite some popular lore to the contrary.

According to the scenario, by the Page time the density of microscopic Einstein-Rosen bridges on the *outer* apparent horizon is about one per Planck area. The back-reaction then would seem to have the potential of modifying the spacetime geometry there strongly enough to eliminate all trapped surfaces. If this happens, the quantum information trapped by the black hole can escape freely, without any complications due to requiring mutual entanglement of the Hawking radiation quanta among themselves. Unitary evolution from a pure state would not require that the Hawking radiation span the entire Hilbert space, with zero von Neumann entropy, after the black hole has disappeared, since it is only the complete system of Hawking quanta and their escaped partners that would have to be in a pure state. Thus my scenario would seem to be consistent with recent arguments of Hawking[24] regarding black hole evaporation and the CPT symmetry expected for a quantum theory of gravity.

There are a number of unsolved issues about how to deal with large quantum back-reaction on the black hole spacetime geometry. The evolution of the quantum state must be thought of as leading to many classical histories, since randomness in the emission of the Hawking radiation over the Page time should lead to macroscopic uncertainty in the position of the black hole. It is certainly not valid to take an expectation value of the quantum energy-momentum tensor with respect to the full quantum state and use it as a source in the classical Einstein equations when the back-reaction is large. Hopefully, the quantum dynamics can be formulated in such a way that fluctuations about a particular well-chosen classical history in a certain region of spacetime can be considered small. Aspects of the issue have been addressed, for instance, by Gell-Mann and Hartle[38] in a cosmological context and by Nomura[39] in connection with black hole evaporation. A fully satisfactory resolution still seems to be out of reach.

Of course, generic black holes are not spherically symmetric, and the interior dynamics of a Kerr black hole would be expected to be quite different than the simple model presented here. A precondition for my scenario of black hole evaporation is that the quantum theory of gravity must resolve spacetime singularities, in the collapse of the star forming the black hole as well as the Schwarzschild singularity. The singularity resolution should not involve a Cauchy horizon, as suggested by some toy models of singularity avoidance in loop quantum gravity[32]. A Cauchy horizon implies a breakdown of unitarity in the sense that there is no unique prediction for the evolution of the quantum state. Also, if trapped surfaces terminate at a Cauchy horizon inside an event horizon, as in a baby universe scenario, all the Hawking partners produced in the entire future evolution of the event horizon would be intersected by the outgoing light-sheet orthogonal to a trapped surface. The number of these could be arbitrarily large, since the event horizon could be prevented from shrinking by continually feeding in enough extra matter and radiation to counter the energy loss from the Hawking radiation.

In an asymptotically AdS spacetime a black hole could be stabilized before the back-reaction is enough to destroy the horizon, if the recycling time for Hawking radiation to be reflected back into the black hole is sustantially less than the Page time. The returning Hawking radiation recaptured by the black hole could keep the entanglement entropy of the black hole from ever getting close to the Bekenstein-Hawking entropy. Such a black hole could persist indefinitely, with the trapped quantum information never reaching the AdS boundary, but the quantum evolution inside the horizon could perhaps still be described by operators in the CFT on the boundary, as suggested by Papadodimas and Raju[40]. In a complicated universe with multiple black holes it seems unlikely that a significant amount of the returning Hawking radiation would be recaptured by the black hole from which it was emitted.

In my picture of black hole evaporation, the Bekenstein-Hawking entropy is strictly a coarse-grained entropy, reflecting complete ignorance about the internal properties of the quantum fields, and is only an upper limit to the actual quantum information contained within the black hole. It is like the entropy per unit area of a Rindler horizon in Minkowski spacetime, which says nothing about the von Neumann entropy of excitations from the Minkowski vacuum behind that horizon. An external observer is not actually completely ignorant of the interior quantum state, since he can know how and from what sort of initial quantum state the black hole was formed. For a black hole formed from gravitational collapse, with quantum fields initially in their vacuum state, the von Neumann entropy starts out small and increases with the number of Hawking partners created during the evaporation process. A large black hole is fundamentally just a region of spacetime, bounded by a hypersurface (the apparent horizon) that can be either spacelike or timelike, but is close to being null for a quiescent large black hole. If the black hole is formed from a pure quantum state, its von Neumann entropy and entanglement entropy should be identical. If the matter and radiation collapsing to form a black hole is not entangled with the outside universe, then it does not contribute to the von Neumann entropy. Most of the von Neumann entropy is just counting the number of Hawking partners and approximately equals the von Neumann entropy of the Hawking radiation emitted. The von Neumann entropy of the black hole keeps increasing as long as trapped surfaces exist, unless the Hawking quanta entangled with the Hawking partners are recycled back into the black hole on a time scale smaller than the Page time.

A full discussion of the proper understanding of black hole entropy in the context of my picture of black hole evaporation is beyond the scope of this paper. A review by Wald[41] discusses many of the issues connected with the thermodynamics of black holes, entropy bounds, etc. A general remark is that large black holes formed by gravitational collapse with quantum fields initially in their vacuum state may not be representative of all possible quantum states of black holes or of all possible thermalized brane/string configurations related to black holes.

If trapped surfaces disappear and the black hole "explodes" approaching the Page time, as suggested here, the observational signals from primordial black holes decaying in the present universe should be quite different from the traditional scenario of evaporation down to the Planck scale[25]. Black holes with a lifetime comparable to the age of the universe, with a mass of about $3 \times 10^{14}$ gm, would not be able to emit particles with energies much larger than tens of MeV, and might be more difficult to detect than black holes emitting much higher energy gamma rays in the final stages of evaporation in the conventional scenario.

ACKNOWLEDGEMENTS

The author thanks Andreas Karch and Ivan Muzinich for a number of very helpful discussions, and Gary Horowitz for not being too negative when I discussed an early stage of these ideas with him and useful comments on a draft of the paper. Also, thanks to the Perimeter Institute for supporting several visits to Waterloo, where stimulating interactions with a number of people helped me appreciate some of the issues involved and where some of the work in writing the paper was done.